\documentclass[a4paper]{jpconf}
\usepackage{graphicx}
\usepackage{subfigure}
\usepackage{ae}
\usepackage{url}
\newcommand{\etab}{\eta_b}
\newcommand{\Bc}{B_c^-}
\newcommand{\BSS}{B^{**}}
\newcommand{\BsSS}{B_s^{**}}
\newcommand{\jpsi}{J/\psi}
\newcommand{\fb}{\textrm{fb}^{-1}}
\newcommand{\GeV}{\textrm{GeV/c}^{2}}
\newcommand{\MeV}{\textrm{MeV/c}^{2}}
\begin{document}
\title{Latest results on b--hadron spectroscopy from CDF}
\author{Andreas Gessler for the CDF Collaboration}
\address{Institut f\"ur Experimentelle Kernphysik, Universit\"at Karlsruhe (TH),
Postfach 6980, 76128 Karlsruhe, Germany}
\ead{gessler@fnal.gov}
\begin{abstract}
B--hadron spectroscopy presents an interesting window for the study of QCD. CDF
has performed a number of studies involving the production and spectroscopy of B
mesons. Among these studies are the first direct observation of the $B_c$, and
the first observation of both narrow states of the $B_s^{**}$. In addition,
measurements are made of the $B^{**}$ masses and widths and the best limit on
the production of $\eta_b$ is set.
\end{abstract}
%
%
\section{Observation of the $\Bc$}
\begin{figure}[h!b]
\subfigure[Mass distribution of the $B^-$ candidates (reference decay).]
{
	\includegraphics[width=0.40\textwidth]{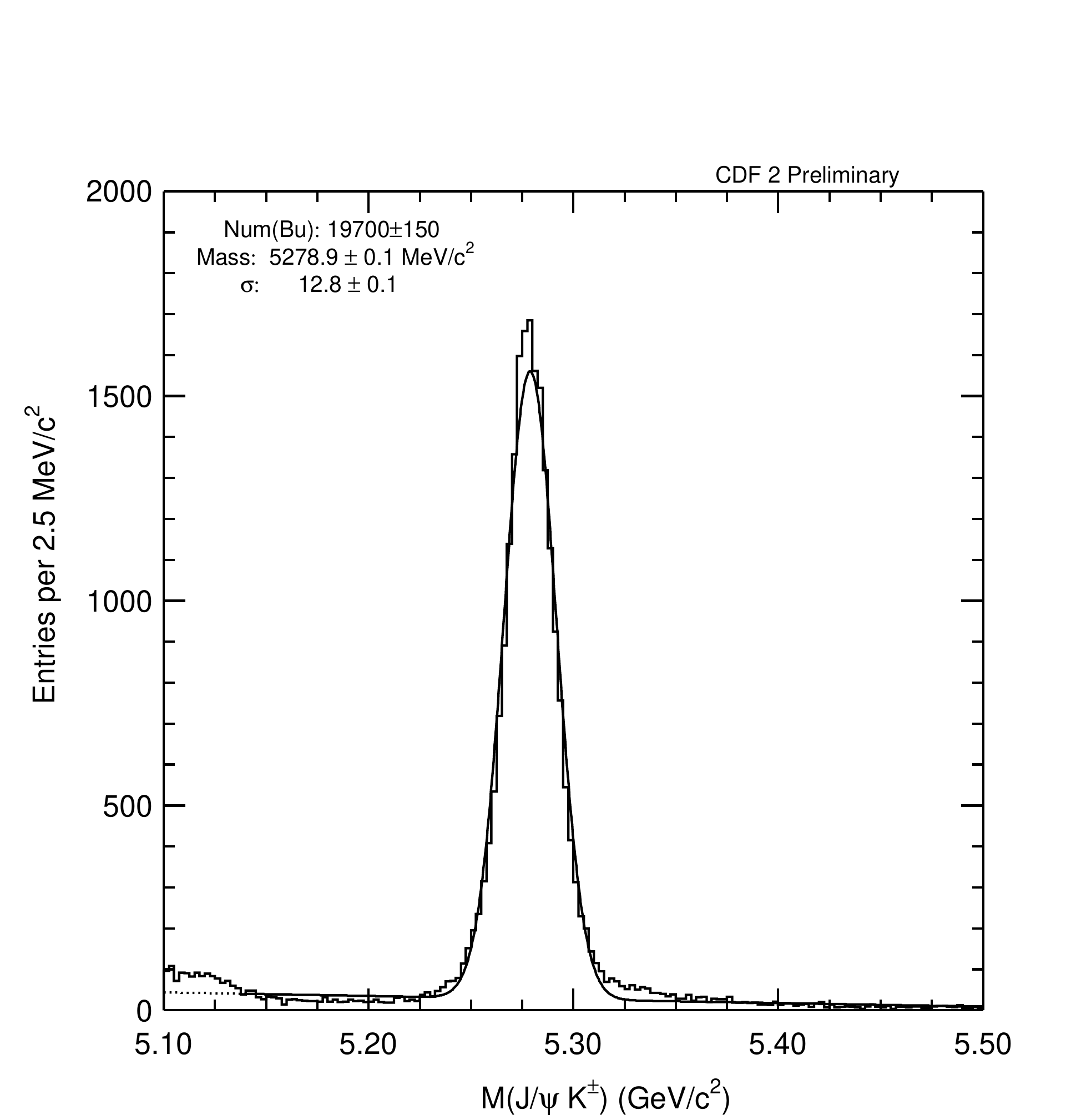}
	\label{fig:BcBuMass}
}
\hfill
\subfigure[Mass distribution of the $\Bc$ candidates.]
{
	\includegraphics[width=0.58\textwidth]{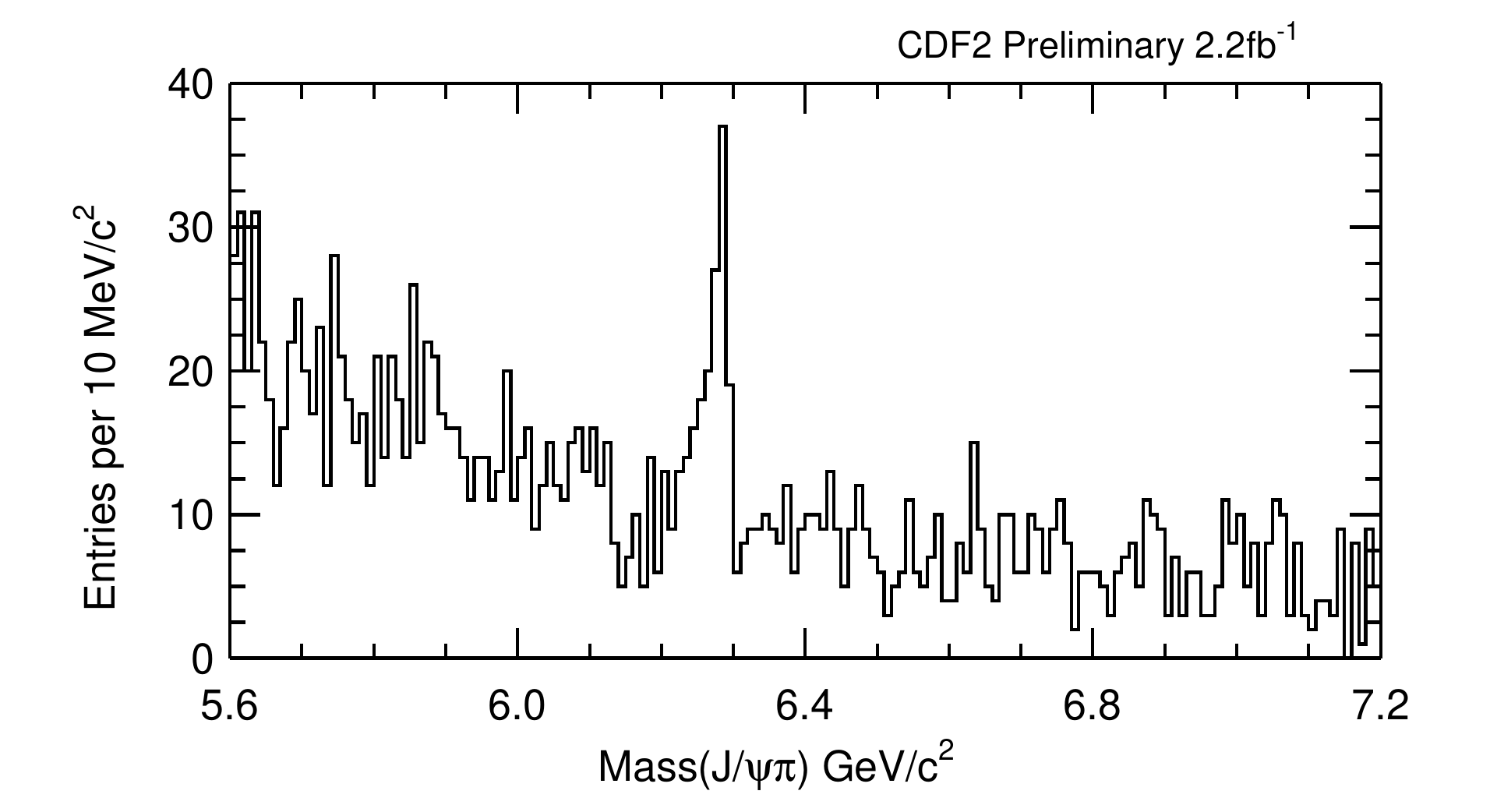}
	\label{fig:BcMass}
}
\caption{Observation of the $\Bc$.\label{fig:Bc}}
\end{figure}

A first evidence of the $\Bc \rightarrow \jpsi\,\pi^-$ was obtained
in~\cite{bsubc}, and after confirmed in higher statistics
samples\footnote{Charge conjugated decays are also implied in all analyses
reported in this article.}~\cite{bsubc11},~\cite{bsubc22}.
The candidate selection is optimized using the reference decay $B^- \rightarrow
\jpsi K^-$ which has similar kinematics but with much higher statistics. 
The $B^-$ selection is applied to $(\jpsi\,\pi^-)$ combinations to search for
$\Bc$ candidates. Figure~\ref{fig:BcBuMass} shows the invariant mass
distribution of the selected candidates on the reference decay and
Figure~\ref{fig:BcMass} shows the invariant mass of the $\Bc$ candidates. To
measure the mass an unbinned maximum likelihood fit is used with a linear and a
Gaussian function for describing the background and the signal, respectively. In
total, there are $87.1\pm12.8$ signal events and the measured mass is
$m\left(\Bc\right) = 6274.1 \pm 3.2 \pm 2.6 \, \MeV$. The determined signal
significance is greater than $8\sigma$.
%
%
%
\section{Search for the $\etab$}
\begin{figure}[h!b]
\subfigure[Mass distribution of the $\etab$ candidates.]
{
	\includegraphics[width=0.45\textwidth]{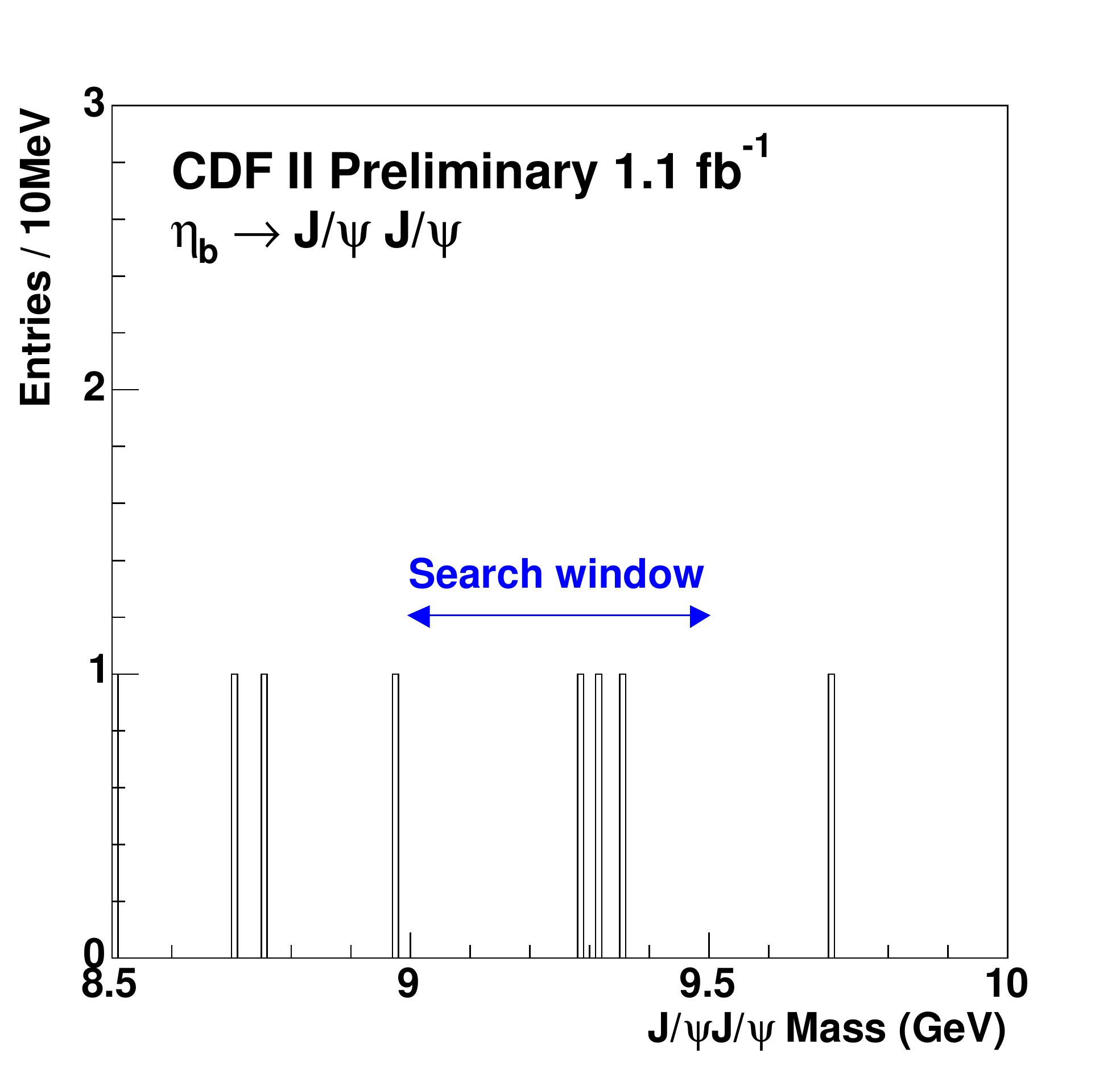}
	\label{fig:etabMass}
}
\hfill
\subfigure[Yield upper limit for $\etab$ search at 95\% C.L.]
{
	\includegraphics[width=0.45\textwidth]{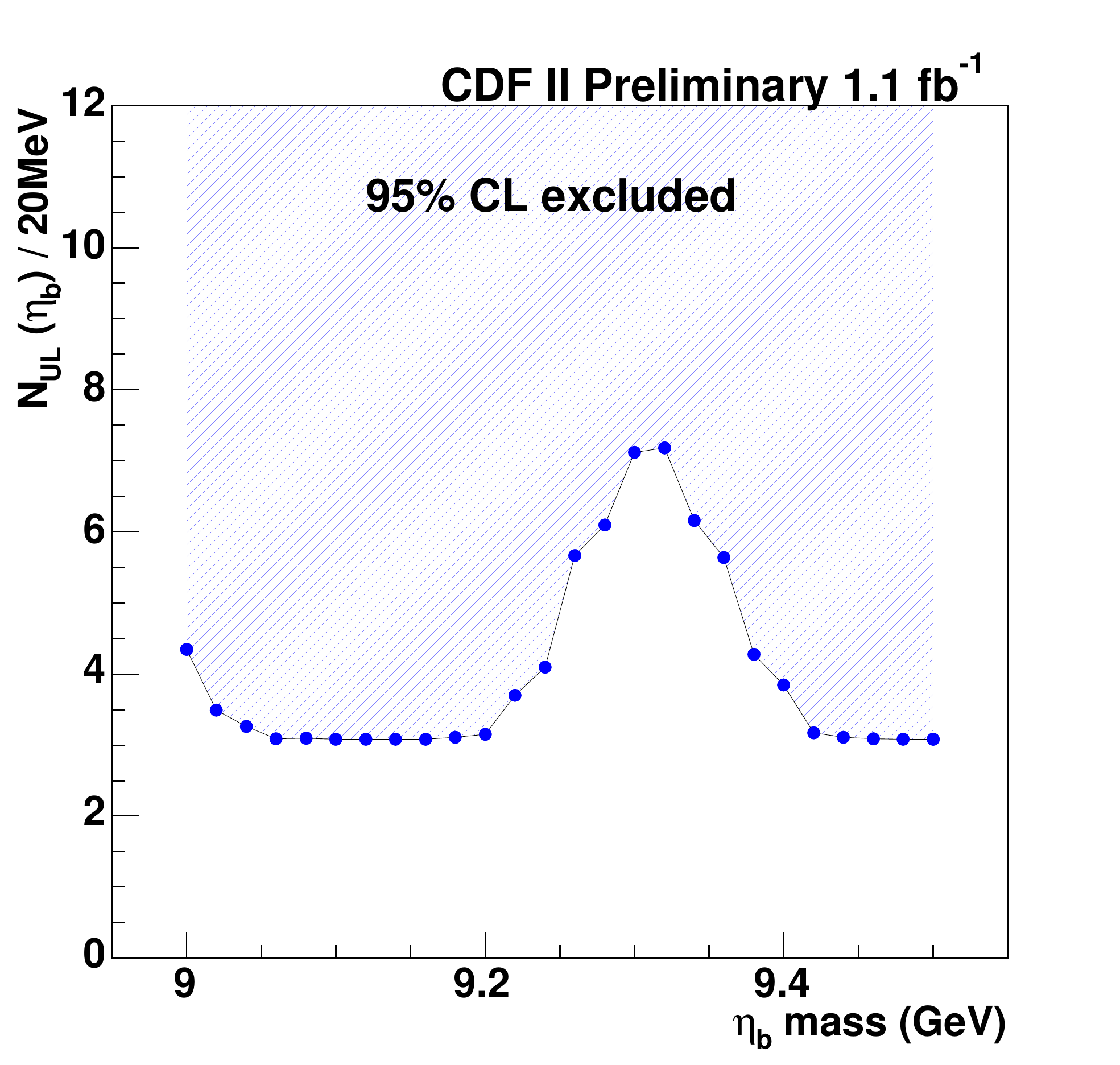}
	\label{fig:etabLimit}
}
\caption{Search for the $\etab$.\label{fig:etab}}
\end{figure}

The $\etab$ is the last undiscovered ground state meson. The search for the
$\etab$ at CDF~\cite{etab} is performed through the exclusive decay into two
$\jpsi$'s. The theoretically predicted branching ratio for this decay is rather
low: BR$\left(\etab\rightarrow\jpsi\jpsi\right) = 7 \cdot 10^{-4\pm1}$.
Together with the calculated differential cross section this corresponds to a
rough estimate of $0.2-20$ expected events within $1.1\,\fb$ of data.\\
Since no significant resonance peak is seen in the mass spectrum, the selection
cuts are tightened to avoid regions where the efficiencies are not well
understood. Thereafter, three events are remaining in the search window which can
be seen in Figure~\ref{fig:etabMass}.
Because only three events are observed, a 95\% C.L. upper limit on the number of
$\etab$ produced is calculated using a Bayesian method. A maximum of $7.2$
events is seen at $9.32\,\GeV$ in the $\etab$ yield upper limit as a function of
$\jpsi$--$\jpsi$ invariant mass (Figure~\ref{fig:etabLimit}). This limit is
translated to an upper limit relative to an inclusive $\jpsi$ production from
b--hadron decays of
$
\sigma\left(p\overline{p}\rightarrow\etab\,X\right) \cdot
BR\left(\etab\rightarrow\jpsi\,\jpsi\right) /
\sigma\left(p\overline{p}\rightarrow H_b\rightarrow\jpsi\,X\right) < 5 \cdot
10^{-3}
$.	
%
%
%
%
\section{Orbitally excited b--mesons}
\begin{figure}[h!b]
\subfigure[Q value distribution of the $\BsSS$ candidates.]
{
	\includegraphics[width=0.45\textwidth]{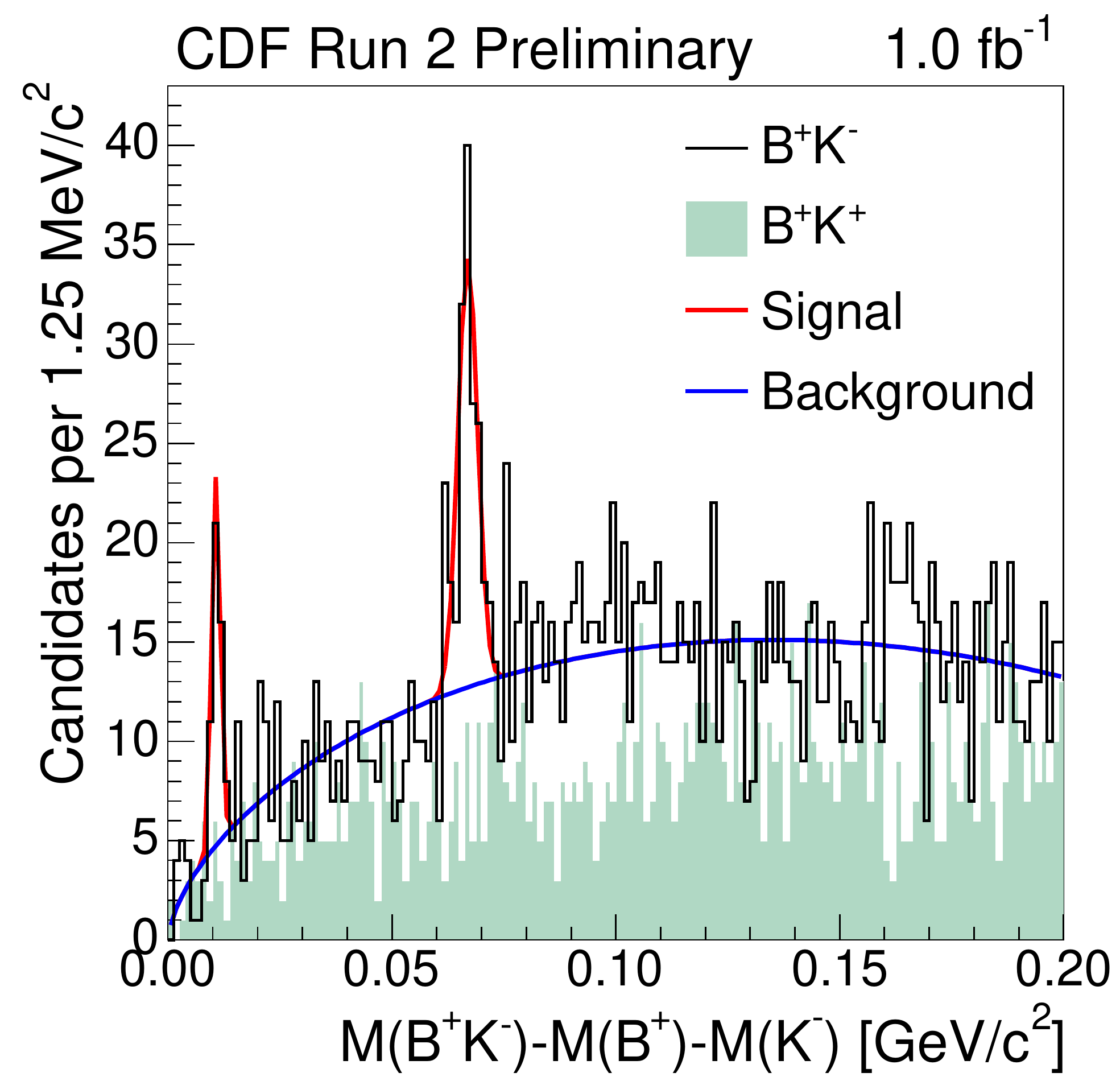}
	\label{fig:BsSSQValue}
}
\hfill
\subfigure[Q value distribution of the $\BSS$ candidates.]
{
	\includegraphics[width=0.45\textwidth]{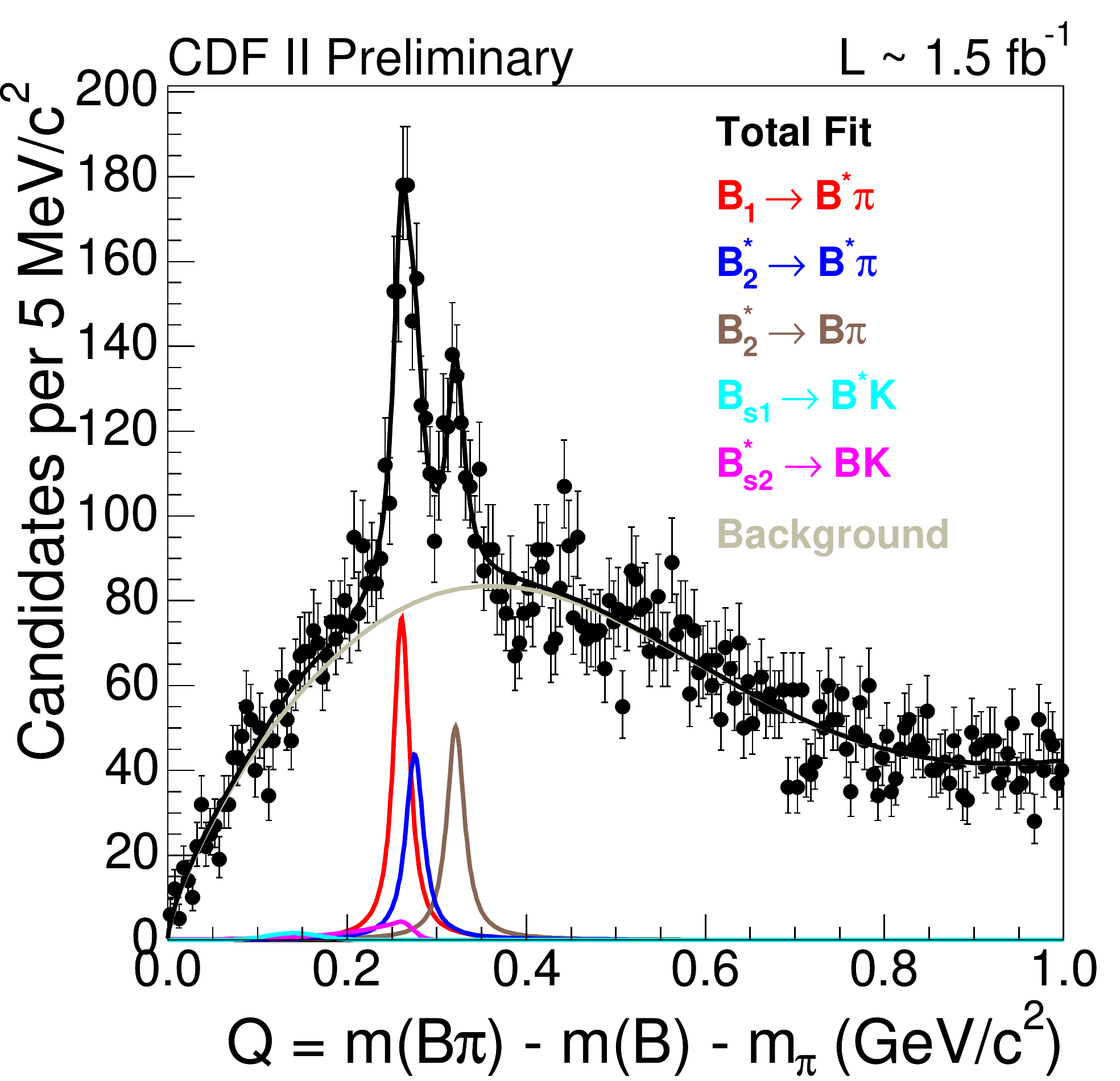}
	\label{fig:BSSQValue}
}
\caption{Orbitally excited b--mesons.\label{fig:BSS}}
\end{figure}

The mass spectra of orbitally excited b--mesons can be described using the Heavy
Quark Effective Theory (HQET). HQET describes mesons consisting of a heavy and a
light quark in the limit where the mass of the heavy quark is considered to be
infinity.\\
In the above limit, the spins of the heavy and light quark are decoupled. Thus,
for the angular momentum $L=1$, the total spin, $\vec{j}_q=\vec{s}_q +\vec{L}$, of
the light quark can either be $j_q=\frac{1}{2}$ or  $j_q= \frac{3}{2}$. Adding
the angular momentum of the light system, $j_q$, with the spin of the heavy
quark, $S_Q$, one obtains four spin states with $J^P=0^+, 1^+, 1^+, 2^+$ which
are collectively called $B_{(s)}^{**}$ states.\\
Orbitally excited states are searched for $B_s$ and $B_d$ mesons~\cite{bsss}.
The excited $B_s$ states $B_{s1}$ and $B_{s2}^*$ (summarized as $B_s^{**}$
states) were observed through their decay to $B^+\,K^-$, with the $B^+$ decaying
into $\jpsi\,K^+$ or $\overline{D}^0\,\pi^+$. \\
The $B^+$ candidates are preselected by using distinct neural networks for each
$B^+$ decay. The selection of the $\BsSS$ candidates is done by using again
different neural networks in each $B^+$ decay mode. For the final selection cuts
on the network output and the number of candidates per event are applied.
Figure~\ref{fig:BsSSQValue} shows the $Q$ value distribution of the $\BsSS$
candidates. The $Q$ value is the defined as
$Q=m\left(B^+K^-\right)-m\left(B^+\right)-m\left(K^-\right)$.\\
To measured the $B_s^{**}$ masses an unbinned maximum likelihood fit is used with
a Gaussian and an exponential function to describe the signal and the
background respectively. The measured masses are $m\left(B_{s2}^*\right) =
5839.6 \pm 0.3 \pm 0.64\,\MeV$ and $m\left(B_{s1}\right) = 5829.4 \pm 0.21 \pm
0.2\,\MeV$. The determined signal significance is larger than $5\sigma$.\\
The analysis of the $\BSS$ mesons is analog to the $\BsSS$ analysis. It is
performed on the decay $B^{**}\rightarrow B^+\,\pi^-$ with the same $B^+$ decay
modes using the same neural networks for preselecting the $B^+$ candidates as
for the $\BsSS$ analysis. Additionally the decay mode
$B^+\rightarrow\overline{D}^0\,\pi^+\pi^+\pi^-$ is also used. The final
candidate selection is based again on distinct neural networks and cuts on the
same quantities as in the $\BsSS$ analysis. Figure~\ref{fig:BSSQValue} shows the
$Q$ distribution of the $\BSS$ candidates. The masses are measured with a
non--relativistic Breit--Wigner distribution modelling the signal. The masses
are $m\left(B_1^0\right) = 5725.3^{+1.6+0.8}_{-2.1-1.1}\,\MeV$ and
$m\left(B_2^{*0}\right) = 5739.9^{+1.7+0.5}_{-1.8-0.6}\,\MeV$ whereas the width
of the $B_2^{*0}$ is measured to be $\Gamma\left(B_2^{*0}\right) =
22.1^{+3.6+3.5}_{-3.1-2.6}\,\MeV$.
%
%
%

%
%
%
\end{document}